\documentclass[twocolumn,floatfix,showpacs,preprintnumbers,amsmath,amssymb]{revtex4-1}
\usepackage{graphicx}

\usepackage{float}
\usepackage{color}
\usepackage{hyperref} 
\hypersetup{
colorlinks,%
citecolor=blue,%
filecolor=black,%
linkcolor=black,%
}
\usepackage{dcolumn}
\usepackage{bm,verbatim}
\usepackage{amsmath} 
\usepackage{bm,lineno} 
\usepackage{natbib} 
\begin{document}
\title{Transient Dynamics in Molecular Junctions:  Coherent Bichromophoric Molecular Electron Pumps}
\author{Roie Volkovich}
\author{Uri Peskin}
\email{uri@tx.technion.ac.il}
\affiliation{Schulich Faculty of Chemistry
and the Lise Meitner Center for Computational Quantum Chemistry\\
Technion-Israel Institute of Technology, Haifa 32000, Israel}
\begin{abstract}

The possibility of using single molecule junctions as electron pumps for energy conversion and storage is considered. It is argued that the small dimensions of these systems enable to make use of unique intra-molecular quantum coherences in order to pump electrons between two leads and to overcome relaxation processes which tend to suppress the pumping efficiency.
In particular, we demonstrate that a selective transient excitation of one chromophore in a bi-chromophoric donor-bridge-acceptor molecular junction model yields currents which transfer charge (electron and holes) unevenly to the two leads in the absence of a bias potential. The utility of this mechanism for charge pumping in steady state conditions is proposed.

\end{abstract}
\maketitle

Single molecule junctions were explored and studied extensively in recent years\cite{123456} primarily due to their potential
for nano-electronics devices. Indeed, non trivial current-voltage characteristics were measured, reflecting the unique properties of organic molecules, associated with their detailed structure\cite{78910111214151617,13,18}. Less attention was given to the possibility of using single molecule junction as energy conversion devices. However, the successful implementation of organic molecules in solar cells\cite{gratzel}, and the quest for energy conversion units on the nanoscale, raise an intriguing challenge of designing particular molecular structures that can function as energy conversion units in a junction architecture.
In this work bi-chromophoric molecules are suggested as appropriate candidates for this application.
Asymmetric Donor-Bridge-Acceptor molecules in which two chromophores are separated by a molecular bridge were suggested in the past as effective molecular current rectifiers \cite{26}. Below we study their function as electron pumps in a single molecule junction. The small dimensions of the proposed systems enables to make use of their unique intra-molecular quantum coherences in order to pump electrons between the leads and to over come relaxation processes which tend to suppress the pumping efficiency.
Current generation by electron pumping is facilitated in the absence of bias by a time-dependent excitation. The possibility of controlling electronic currents in molecular junctions by periodic external fields was discussed intensively\cite{18,19,20,21,22}. In particular, monochromatic, bi-chromatic\cite{18,22} or pulsed \cite{23,24,25} electromagnetic radiation fields were proposed as means of controlling the current intensity under steady state conditions. Given an appropriate
asymmetry of the junction such external fields can induce net current generation even in the absence of bias. This was anticipated, e.g, when the coupling to the electrodes differs in the ground and excited molecular state\cite{50}.
Although pumps are usually discussed in the context of periodic driving \cite{27,28,29}, in order to pinpoint the underlying mechanism of the molecular energy converter, we consider below a single pumping cycle in which one of the chromophores (termed "donor") is subjected to a sudden excitation.

\begin{figure} [b]
  \includegraphics [scale=0.4]{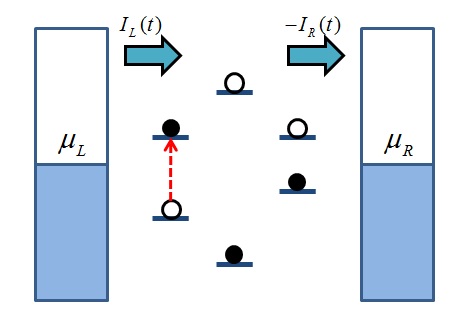}
  \caption{A scheme for DBA molecular junction at zero bias. Notice that the LUMO levels of the D and A sites are degenerate but the HOMO-LUMO energy gaps are different. An initial D excitation (marked by the dashed arrow) is followed by transient electronic currents (thick arrows) from left to right. }\label{Fig.1}
\end{figure}

The DBA molecule is represented as three electronic sites corresponding to the donor(D), bridge(B) and acceptor(A) molecular groups (see Fig. \ref{Fig.1}). Invoking for simplicity a non interacting spinless electrons picture, the molecular model Hamiltonian reads,
\begin{eqnarray}\label{2.1}
  \hat{H}_M=\sum_{l=1}^{2}[\sum_{m=1}^{3}\varepsilon_{m,l}d_{m,l}^{\dagger}d_{m,l}
  +\sum_{m=2}^{3}(\beta_{m,l}d_{m,l}^{\dagger}d_{m-1,l}+\emph{h.c})] \nonumber\\
\end{eqnarray}
$d^{\dagger}_{m,l}(d_{m,l})$ is a creation (annihilation) operator for an electron at the \emph{l} th single particle state (orbital) of the \emph{m} th molecular site, $\varepsilon_{m,l}$  is the corresponding orbital energy, and $\{\beta_{m,l,l^{'}}\}$ are the inter-site coupling (hopping) integrals between the two manifolds of orbitals at the \emph{m} th and \emph{m-1} th neighboring sites. The indexes $l=1$  and $l=2$  point to the Highest Occupied Molecular Orbital(HOMO) and Lowest Unoccupied Molecular Orbital(LUMO) at each site, and the indexes $m=1,2,3$   represent the D, B and A sites respectively. In view of the relatively large HOMO-LUMO energy gaps, the inter-site couplings were restricted to a diagonal form, $\beta_{m,l,l^{'}}=\beta_{m,l}\delta_{l,l^{'}}$ (off-diagonal terms yielded insignificant changes to the results reported below), and all inter-site couplings were given the same value, $\beta_{m,l}\equiv\beta$. The value of $\beta$ is strongly sensitive to the bridge length and composition, and determines the time scale for coherent tunneling between D and A. Since coherent tunneling plays an important role here, the value was taken larger than
$K_{B}T$.
Asymmetry of the bichromophoric molecule is reflected in different HOMO-LUMO gaps at the donor and acceptor sites, \emph{i.e.}, $\varepsilon_{1,2}-\varepsilon_{1,1}>\varepsilon_{3,2}-\varepsilon_{3,1}$, where the effective electronic coupling between the D and A sites is mediated by the bridge HOMO-LUMO gap. Here we focus on a particular system in which the D and A LUMO states are degenerate, \emph{i.e} $\varepsilon_{1,2}=\varepsilon_{3,2}$. (a similar effect would be obtained if the degeneracy would apply to the HOMO states). A  schematic representation of the model is given in Fig.\ref{Fig.1}. In view of the differences between the HOMO-LUMO gaps at the different molecular sites, direct intra-molecular exciton energy transfer \cite{32} is inefficient, and terms of the type ($\propto d_{m,l}^{\dagger}d_{m,l^{'}}d_{m^{'},l^{'}}^{\dagger}d_{m^{'},l}$) were excluded in Eq.\ref{2.1}.

The terminal sites (D and A) are coupled to two (left and right, respectively) electrodes, so that the full Hamiltonian reads,
\begin{eqnarray}\label{2.2}
  \hat{H}&=&\hat{H}_M+\hat{H}_e+\hat{H}_{e,M}.
\end{eqnarray}
The electrodes are modeled as reservoirs of non interacting electrons,
$\hat{H}_e=\sum_{J=R,L}\sum_{k}\varepsilon_{k,j}b^{\dagger}_{k,j}b_{k,j}$
where $b^{\dagger}_{k,j}(b_{k,j})$ is a creation (annihilation) operator of an electron in the $k$th single particle state of the $J$th electrode' conduction band. The molecule-electrodes coupling terms correspond to electron transfer (hopping) between the D and A sites and the respective electrode, \emph{i.e.},
  $\hat{H}_{e,M}=\hat{V}_{L}\otimes\hat{U}_{L}+\hat{V}_{R}\otimes\hat{U}_{R}+h.c$,
with $\hat{V}_{L}\equiv\sum_{l=1}^{2}{d^{\dagger}_{1,l}}$, $\hat{V}_{R}\equiv\sum_{l=1}^{2}{d^{\dagger}_{3,l}}$, $\hat{U}_L\equiv\sum_{k}{u_{k,L}b_{k,L}}$, and $\hat{U}_R\equiv\sum_{k}{u_{k,R}b_{k,R}}$.
 For simplicity, electron energy transfer from the molecular excitons to electron-hole pairs creation at the leads \cite{33} is not accounted for. The values of $\{u_{k,J}\}$ reflect the spectral density function of the Jth electrode. Below, each electrode is modeled as a half-filled tight binding chain with a bandwidth $|4\gamma_{J}|$ and an electrochemical potential, $\mu_J$. The electrode band is discretized \cite{31}, where the single particle energies are $\varepsilon_{k,j}=\mu_{J}-2|\gamma_{J}|cos[\frac{k\pi}{N+1}]$, and therefore, $u_{k,J}=\xi_{J}\sqrt{\frac{2}{N+1}}sin[\frac{k\pi}{N+1}]$. $\xi_{J}$  is the hopping parameter between the terminal molecule and electrode sites.
\begin{table}
\caption{\label{Table I}Model Parameters (Energy values are in eV)}
\begin{ruledtabular}
\begin{tabular}{*{5}{c}}
$\mu_L,\mu_R$ & $\xi_L,\xi_R$ & $K_{B}T$  & $\gamma_L,\gamma_R$ & $\beta$ \\
\hline
-0.2 & -0.03 & 0.001 & -1 & -0.01 \\
\end{tabular}
\end{ruledtabular}
\vskip 0.1 cm
\begin{ruledtabular}
\begin{tabular}{*{5}{c}}
Fig.& $\left(
     \begin{array}{ccc}
        \varepsilon_{1,2} & \varepsilon_{2,2} & \varepsilon_{3,2} \\
        \varepsilon_{1,1} & \varepsilon_{2,1} & \varepsilon_{3,1} \\
      \end{array}
    \right)$
& $\hat{\rho}_M(0)$& $\Omega$ \\
\hline
2 & $\left(
     \begin{array}{ccc}
        0 & \varepsilon_{2,2} & 0 \\
        -0.3 & -0.6 & -0.25 \\
      \end{array}
    \right)$
&$\left(
    \begin{array}{ccc}
      d_{1,1}d_{1,1}^{\dagger}\otimes d_{1,2}^{\dagger}d_{1,2}  \\
      \otimes d_{2,1}^{\dagger}d_{2,1}\otimes d_{2,2}d_{2,2}^{\dagger} \\
      \otimes d_{3,1}^{\dagger}d_{3,1}\otimes d_{3,2}d_{3,2}^{\dagger}\\
    \end{array}
  \right)$
& N/A \\
3 & $\left(
     \begin{array}{ccc}
        0 & 0.05 & 0 \\
        -0.3 & N/A & N/A \\
      \end{array}
    \right)$
&$\left(
    \begin{array}{c}
      d_{1,1}d_{1,1}^{\dagger}\otimes d_{1,2}^{\dagger}d_{1,2} \\
      \otimes d_{2,2}d_{2,2}^{\dagger}\otimes d_{3,2}d_{3,2}^{\dagger} \\

    \end{array}
  \right)$
& 0.06 \\
\end{tabular}
\end{ruledtabular}
\label{table}
\end{table}

An initial sudden preparation of an exciton at the molecular donor site is represented by a factorized density operator, $\hat{\rho}(0)=\hat{\rho}_{M}(0)\otimes\hat{\rho}_{R}\otimes\hat{\rho}_{L}\cdot \hat{\rho}_{M}(0)$
 is the molecular density which accounts for the population of electrons ($n_{m,l}=1$) or holes ($n_{m,l}=0$) in the molecular single particle states (see Fig.\ref{Fig.1} and Table \ref{Table I}), $\hat{\rho}_{M}(0)=\prod_{\{m,l\}}\hat{\rho}_{m,l}(0)$,
\begin{eqnarray}\label{3.2}
 \hat{\rho}_{m,l}(0)=\left\{
  \begin{array}{c}
    d_{m,l}d_{m,l}^{\dagger}; n_{m,l}=0 \\
    d_{m,l}^{\dagger}d_{m,l}; n_{m,l}=1 \\
  \end{array}
\right.
\end{eqnarray}

The reservoirs are associated with equilibrium densities,
$\hat{\rho}_J=e^{-\frac{1}{K_{B}T}\sum_{k}(\varepsilon_{k,J}-\mu_{J})b_{k,J}^{\dagger}b_{k,j}}/tr[e^{-\frac{1}{K_{B}T}\sum_{k}(\varepsilon_{k,J}-\mu_{J})b_{k,J}^{\dagger}b_{k,j}}]$.
For weak coupling between the molecular system and the electrodes, it is convenient to regards the reservoirs as external baths, and follow the system time-evolution using a Redfield type equation of motion for a reduced system density \cite{30}, $\hat{\rho}_M(t)\equiv tr[\hat{\rho}(t)]$, keeping terms up to second order in $\hat{H}_{e,M}$\cite{31},
\begin{eqnarray} \label{3.4}
  &&\frac{d}{dt}\hat{\rho}_{M}(t)=-\frac{i}{\hbar}[\hat{H}_{M},\hat{\rho}_{M}(t)] \\ &&+\sum_{J=L,R}[\{\hat{F}_{J}(t)\hat{\rho}_M(t)-\hat{\rho}_{M}(t)\hat{\tilde{F}}_{J}^{\dagger}(t)\},\hat{V}_J]+h.c \nonumber
\end{eqnarray}
where, $\hat{F}_{J}(t)=\frac{1}{\hbar^2}\int_0^t{d\tau C_J(\tau)e^{-\frac{i}{\hbar}\hat{H}_M \tau}\hat{V}_{J}^{\dagger}e^{\frac{i}{\hbar}\hat{H}_M \tau}}$
and $\hat{\tilde{F}}_{J}(t)=\frac{1}{\hbar^2}\int_0^t{d\tau \tilde{C}_J(\tau)e^{-\frac{i}{\hbar}\hat{H}_M \tau}\hat{V}_{J}e^{\frac{i}{\hbar}\hat{H}_M \tau}}$, depend on electrode correlation functions,
$C_J(\tau)=\sum_k{|u_{k,J}|^2  e^{-\frac{i}{\hbar}\varepsilon_{k,J}\tau}[1-f_J(\varepsilon_{k,J})]}$
and $\tilde{C}_J(\tau)=\sum_k{|u_{k,J}|^2 e^{\frac{i}{\hbar}\varepsilon_{k,J}\tau}f_J(\varepsilon_{k,J})}$.
$f_J(\varepsilon_{k,J})\equiv\frac{1}{1+e^{(\varepsilon_{k,J}-\mu_J)/(K_B T)}}$, is the electrode Fermi function.
The change in the charge on the molecular system is defined as,
$\frac{dQ_{M}}{dt}=e\cdot tr[\hat{N}\frac{d}{dt}\hat{\rho}_{M}(t)]$,
where $\hat{N}_M=\sum_{m=1}^3\sum_{l=1}^2{d_{m,l}^{\dagger}d_{m,l}}$ is the electronic number operator in the system. Using $[\hat{N},\hat{H}_M]=0$,
one can express the charge dynamics in terms of additive contributions from the two electrodes,
\emph{i.e.},
$\frac{dQ_M}{dt}\equiv I_R (t)+I_L (t)$,
where the transient current from the electrode into the molecule read,
\begin{eqnarray} \label{3.15}
I_J (t)= 4eRe (tr\{\hat{N} [\{\hat{F}_J (t)\hat{\rho}_M (t)-\hat{\rho}_M (t) \hat{\tilde{F}}_J^{\dagger} (t)\},\hat{V}_J]\}) \qquad
\end{eqnarray}
The integral, $Q_{J}=\int_0^{\infty}{I_{J}(t)dt}$, is the accumulated charge due to electron and hole transfer (positive values correspond to electron transfer into the molecule), which depend on the intramolecular dynamics, including intramolecular coherence.
For the DBA molecule illustrated in Fig.\ref{Fig.1}, the HOMO energies at the D and A sites are detuned, such that the effective electronic coupling between them is week regardless of the B HOMO energy. In contrast, the two LUMO energies are resonant and therefore the electronic coupling between them depends sensitively on the B LUMO energy.
\begin{figure} [tR!]
  \includegraphics [scale=0.4]{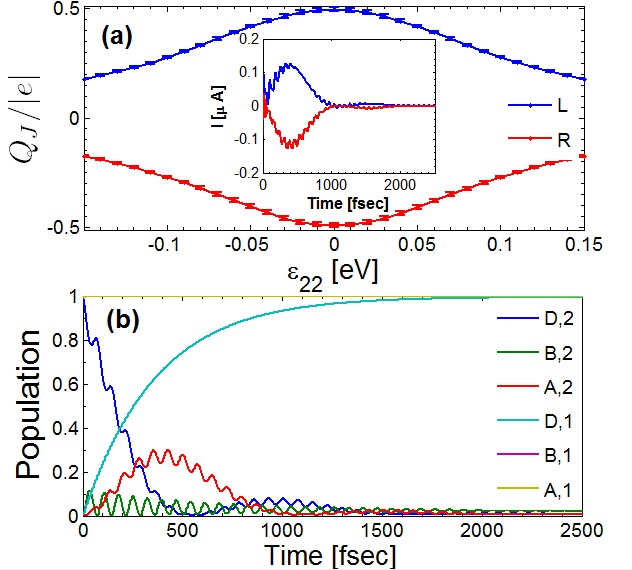}
  \caption{(a) Accumulated charges at the electrodes (current integrals) as functions of the bridge LUMO energy. Inset: the transient currents for a selected energy, $\varepsilon_{2,2}=0.05 eV$.
(b) Transient populations of the molecular states following a sudden excitation of the donor for $\varepsilon_{2,2}=0.05eV$. Each molecular orbital  is denoted by its site (D, B or A) and energy indexes (1 for HOMO and 2 for LUMO).}\label{Fig.2}
\end{figure}

In Fig.\ref{Fig.2} the accumulated charges are plotted as functions of the bridge LUMO energy, $\varepsilon_{2,2}$ \cite{34} (the model parameters are detailed in Table \ref{Table I} ). As one can see, there is a net charge transfer from the left electrode to the right one, $Q_L>0,Q_R<0$, (see also the inset, for the transient currents). The "charge pumping" efficiency, regarded here as the net amount of charge transferred from the left electrode to the right one, increases and reaches a maximum as the B level approaches resonance with the D and A levels.
The charge transfer between the electrodes reflects intramolecular dynamics as presented in Fig.\ref{Fig.2}b. The apparent dominant processes following the initial excitation at the donor are a decay of the hole at the donor HOMO (see D,1), and two decay processes of the electron from the donor LUMO: One is a direct relaxation to the left electrode (see D,2) and the other is an indirect relaxation to the right electrode following coherent tunneling from the D to the A LUMO level (see A,2). When the bridge LUMO is detuned from the D/A LUMOs (\emph{i.e.}, when $|\beta|<<|\varepsilon_{2,2}-\varepsilon_{1,2}|$) the period of the DA tunneling oscillations can be approximated by the McConnell formula \cite{35}, $\tau_{DA}\propto \frac{\hbar|\varepsilon_{2,2}-\varepsilon_{1,2}|}{|\beta|^2}$, and becomes shorter as the gap between the B and the D(A) orbital energies decreases. As the period becomes sufficiently short on the time scale of the electron relaxation to the leads, electron hopping to the right becomes as frequent as electron hopping to the left. Since the hole relaxes predominantly to the left lead, the balance of these processes is net transfer of negative charge from left to right.

It is emphasized that although the simple model outlined above is formulated in the language of
non interacting particles, it can guide a search for real bichromophoric systems where the proposed mechanism of coherent pumping following
a local excitation of a donor chromophore should be observed. Considering a bichromophoric molecule where the two different chromophores,
are weakly coupled to each other and to the leads, the condition for near degeneracy of the two LUMOs in the
non-interacting model corresponds to a generalized condition for a near degeneracy between the first excited (many body) state
of one of the chromophores the "Donor" and a charge transfer (many body) state of the bichromophoric system. In such a case, local excitation of the Donor
would result in to intra-molecular dynamical charge oscillations, followed by the proposed charge pumping.

\begin{figure} 
  \includegraphics [scale=0.3]{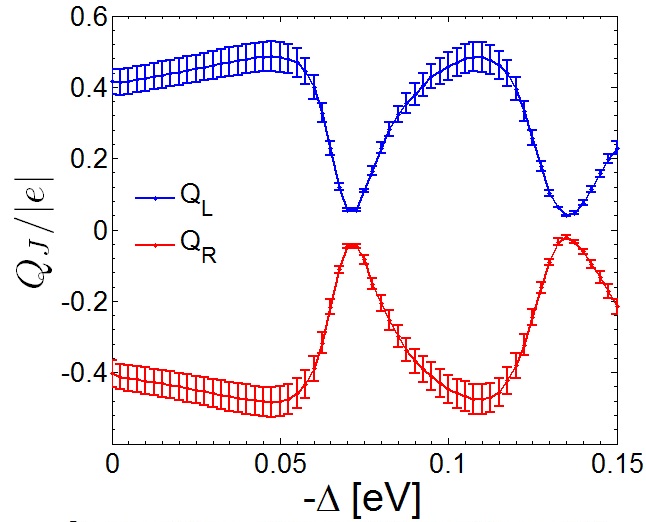}
  \caption{Accumulated charges at the electrodes (current integrals) as functions of the bridge reorganization energy. }\label{Fig.4}
\end{figure}

Let us now consider a possible effect of bridge vibrations on the pumping efficiency. For this purpose we extend the model to account for linear on-site coupling of the bridge excited state to a harmonic mode.  The model DBA Hamiltonian (Eq.(1)) is replaced by,
\begin{eqnarray}\label{5.1}
  \hat{H}_{M,vib} = \hat{H}_{M}+ \frac{\lambda}{\sqrt{2}}(c^{\dagger}+c)d_{2,2}^{\dagger}d_{2,2}+\hbar\Omega({c^{\dagger}c+1/2})\qquad
  \end{eqnarray}
The last term is the nuclear contribution, where $c^{\dagger}$ ($c$) is a creation (annihilation) operator of a vibration quantum, $\hbar\Omega$, and the vibronic coupling strength is measured by the bridge reorganization energy, $\Delta=\frac{\lambda^2}{2\hbar \Omega}$.
In Fig.\ref{Fig.4}, the effect of $\Delta$ on the pumping efficiency is demonstrated. The acceptor and bridge HOMO states, which remain occupied and hardly participate in the dynamics (See Fig.\ref{Fig.2}), were excluded in this case (see Table \ref{Table I}), and the initial electronic excitation was represented as $\hat{\rho}_{M,vib}(0)=\hat{\rho}_{M}(0)\otimes|0\rangle\langle 0|$, where ($|0\rangle$) is the vibrational ground state.
The left-to-right electron pumping efficiency is shown to modulate as a function of the vibronic coupling strength, where efficiency peaks ($-Q_R=Q_L\approx0.5$) are obtained near $\Delta\approx0.05,0.11$. As above, the increased efficiency corresponds to resonances between the donor and acceptor LUMO orbitals and, in this case, {\it vibronic} eigenstates of the bridge, where the peak efficiency corresponds to $\varepsilon_{2,2}-\Delta\approx0,\hbar\Omega$  respectively. Again, approaching the resonance condition guarantees that the frequency of tunneling oscillations between the D and A exceeds the rate of charge relaxation to the electrodes, and thus the charge pumping efficiency increases.
Notice that the efficiency peaks in Fig.\ref{Fig.4} are characterized by a slow rise, followed by a fast fall as $\Delta$ increases. This reflects a characteristic Franck-Condon blockade effect \cite{13}. Focusing on the first resonance, one can approximate the tunneling frequency by a generalized McConnell formula \cite{36}, $\nu=\frac{2\beta^2 e^{-\frac{\Delta}{\hbar\Omega}}}{\hbar|\varepsilon_{2,2}-\Delta|}$, (for $\varepsilon_{1,2}=\varepsilon_{3,2}=0$). As $\Delta$ rises from zero, two effects are competing. The energy barrier for through-bridge tunneling ($|\varepsilon_{2,2}-\Delta|$) decreases, and the hopping frequency to the bridge, scaled by the Franck Condon (FC) overlap between the ground vibrational states of the empty and electronically charged bridge, decreases as well. For sufficiently small $\Delta$ the exponential drop in the FC factor is minor, and the decrease in the barrier energy leads to increasing tunneling frequency. Increasing $\Delta$ farther, beyond the resonance point, the tunneling frequency decreases due to a rise in $|\varepsilon_{2,2}-\Delta|$, but the fall is much sharper than the rise since the donor-bridge hoping is additionally blocked as the FC factor becomes exponentially small. A similar qualitative modulation is observed for larger $\Delta$ values due to the first excited vibration state of the bridge.

In conclusion, the versatility of organic compounds and the ability to control their electronic structure by chemical substitutions should make the preparation of bi-chromophoric DBA molecules as discussed here a realistic task. The advances in fabrication of molecular junctions (of single molecules or of ordered monolayers) suggest that devices can be based on the single molecule properties. In particular, irradiation of one of the chromophores can be converted into charge pumping even in the absence of a bias potential.
The focus of the present work was on a single pumping cycle. An extension of this study to account for the details of the excitation field would enable to consider the operation of a multi-cycle (periodic) electron pump at steady state. According to the above analysis, the pumping direction would reflect the selection of either one of the two chromophores as the "donor" by the excitation field. Moreover, the possibility for pumping at steady state against an
opposing bias voltage suggests the utility of the proposed molecular junction for electric energy storage.
The analysis of the appropriate field parameters that would induce the required functions given the particular molecular electronic structure is a future challenge, as well as the inclusion of several effects that were left out for simplicity in the present work, but are not expected to change its qualitative conclusions. These include the role of Coulomb repulsion and electronic correlations at the molecule, the vibronic structure of the chromophores, and the effect of energy transfer to the leads. This research was supported by the Israel-U.S. Bi-national Science Foundation and by the Israel Science Foundation.


\begin{thebibliography}{99}
\bibitem{34}The error bars were set according to the artificial residual current at asymptotic time. This current reflects asymmetry in the error of the Redfield calculation between the two electrodes.
\end{thebibliography}
\end{document}